\documentclass[showpacs,showkeys,twocolumn,preprintnumbers,nofootinbib,groupedaddress,superscriptaddress,amsmath,amssymb,showabstract]{revtex4-1}

\usepackage{graphicx}
\usepackage{dcolumn} 
\usepackage{bm}
\usepackage{amssymb}
\usepackage{amsmath}
\usepackage{epsfig}    
\usepackage{color}
\usepackage{slashed}
\usepackage{hyperref} 

\begin{document} 

\title{Multi-photon signatures as a probe of CP-violation in extended Higgs sectors}

\preprint{OU-HET-1197}

\author{Shinya Kanemura}
\affiliation{Department of Physics, Osaka University, Toyonaka, Osaka 560-0043, Japan}

\author{Kento Katayama}
\affiliation{Department of Physics, Osaka University, Toyonaka, Osaka 560-0043, Japan}

\author{Tanmoy Mondal}
\affiliation{Department of Physics, Osaka University, Toyonaka, Osaka 560-0043, Japan}
\affiliation{Birla Institute of Technology and Science, Pilani, 333031, Rajasthan, India}

\author{Kei Yagyu}
\affiliation{Department of Physics, Osaka University, Toyonaka, Osaka 560-0043, Japan}

\begin{abstract}

We propose a novel signature with four-photon final states to probe CP-violating (CPV) extended Higgs sectors via $f\bar{f} \to Z^* \to H_1H_2 \to 4\gamma$ processes with $H_{1,2}$ being additional neutral Higgs bosons. 
We focus on the nearly Higgs alignment scenario, in which the discovered Higgs boson almost corresponds to a neutral scalar state
belonging to the isospin doublet field with the vacuum expectation value $v \simeq 246$ GeV. 
We show that the branching ratios of $H_{1,2} \to \gamma\gamma$ can simultaneously be sizable when CPV phases in the Higgs potential are of order one due to the enhancement of charged-Higgs boson loops.
Such branching ratios can be especially significant when the fermiophobic scenario is taken into account. 
As a simple example, we consider the general two Higgs doublet model, and demonstrate that the cross section for the four-photon process can be 0.1 fb 
at LHC with the masses of $H_{1,2}$ to be a few 100 GeV in the Higgs alignment limit 
under the constraints from electric dipole moments (EDMs) and LHC Run-II data. 
We also illustrate that the searches for EDMs and di-photon resonances at high-luminosity LHC play complementary roles to explore CPV extended Higgs sectors. 

\end{abstract}
\maketitle

\noindent
{\it Introduction --} CP-violation (CPV) is one of the necessary ingredients to explain the baryon asymmetry of the Universe~\cite{Sakharov:1967dj}. 
Although non-zero CPV appears from the Kobayashi-Maskawa phase in the standard model (SM), its amount has been known to be too small to accommodate
the observed value of the baryon asymmetry~\cite{Shaposhnikov:1987tw}. 
Therefore, new physics beyond the SM is required to provide additional sources of CPV. 

A Higgs boson was discovered at LHC in 2012, and its properties, e.g., 
the mass, width and couplings, have been measured from various production and decay channels. 
So far, the observed properties are consistent with those of the Higgs boson in the SM within the theoretical and experimental uncertainties~\cite{ATLAS:2019nkf,CMS:2018uag}. 
This, however, does not necessarily mean that the Higgs sector is the minimal one assumed in the SM. 
In fact, it is indeed possible to realize non-minimal Higgs sectors with 
nearly Higgs alignment~\cite{Davidson:2005cw}, in which couplings of the discovered Higgs boson take almost the same values as those of the SM Higgs boson at tree level. 
Since the Higgs alignment can be compatible with CPV in extended Higgs sectors, e.g., in models with multi-Higgs doublets~\cite{Kanemura:2020ibp},  
it is now quite important to investigate CP-violating non-minimal Higgs sectors with the nearly Higgs alignment~\cite{Enomoto:2021dkl,Enomoto:2022rrl,Kanemura:2023juv}. 

Searches for electric dipole moments (EDMs) can provide evidence for CPV in the Higgs sector. 
The EDM experiments severely constrain a possible parameter space in non-minimal Higgs sectors with CPV, 
and might be able to test such a Higgs sector in future experiments. 
In particular, their sensitivities have been significantly improved after the Higgs boson discovery. 
For instance, the magnitude of the electron EDM (eEDM) has been constrained to be smaller than $4.1\times 10^{-30}e\,\text{cm}$ (90\% CL)~\cite{Roussy:2022cmp}. 
%
In addition to the EDMs, CP-violating effects can be tested at high energy collider experiments. 
It has been known that the decay of neutral Higgs bosons 
into a tau-pair can be used to extract the CP-violating phase from the difference of the azimuthal angles defined by the tau decay plane~\cite{Kuhn:1982di,Grzadkowski:1995rx,Hagiwara:2012vz}, and 
the possibility of measuring the phase has been discussed at LHC in Refs.~\cite{Harnik:2013aja,Berge:2008dr,Berge:2015nua,Dolan:2014upa,Przedzinski:2014pla} and at future electron-positron colliders in Refs.~\cite{Jeans:2018anq,Kanemura:2021atq}. 
The CP nature of the neutral Higgs boson can also be extracted via the top Yukawa coupling~\cite{Boudjema:2015nda,Faroughy:2019ird,Cheung:2020ugr}, the diboson decay~\cite{Keus:2015hva} 
and also from Higgs to Higgs decays~\cite{Low:2020iua}. 

In this Letter, we propose a novel approach to test non-minimal Higgs sectors with CPV at collider experiments. 
We focus on the four-photon final state driven by the electroweak (EW) pair production of additional neutral Higgs bosons $H_1$ and $H_2$ (the discovered Higgs boson with the mass of 125 GeV is denoted as $h$) 
with their subsequent di-photon decays: 
\begin{align}
f\bar{f} \to Z^* \to H_1H_2\to 4\gamma. \label{eq:4photon}
\end{align}
We show that the cross section for the above process can be significant in the presence of charged Higgs bosons when 
the CP-violating phase in the Higgs potential is sizable. 
We would like to emphasize that our approach can be applied to a plethora of extended Higgs sectors with CPV, and offers robust probe of CPV in the Higgs potential
since the production part $f\bar{f} \to H_1H_2$ is purely determined by the gauge coupling, by which the CP-violating nature can be extracted from the decays of $H_{1,2}$. 

{\it General setup --} Let us first consider a rather general setup in the EW $SU(2)_I\times U(1)_Y$ gauge theory with extended Higgs sectors. 
We then discuss concrete models later. 

Suppose that $\Phi$ and $\varphi$ are respectively the isospin Higgs doublet with the hypercharge 1/2 and a complex scalar multiplet with the hypercharge $Y_\varphi$ containing a neutral component $\varphi^0$. 
We focus on the nearly Higgs alignment scenario as it is favored by the current LHC data~\cite{ATLAS:2019nkf,CMS:2018uag}, 
where the Fermi constant $G_F$ is mainly given by the vacuum expectation value (VEV) $v$, i.e., $v \equiv \sqrt{2}\langle \Phi^0\rangle \simeq (\sqrt{2}G_F)^{-1/2}$ and $h \equiv \sqrt{2}\Re\Phi^0 - v$ is supposed to be almost the mass eigenstate. 
In the following, we first consider the case with the exact Higgs alignment, and then discuss the consequence of a slight deviation from the alignment limit.
The real part $\varphi_H \equiv \sqrt{2}\Re(\varphi^0)$ and the imaginary part $\varphi_A \equiv \sqrt{2}\Im(\varphi^0)$ can mix if the Higgs potential contains CP-violating phases. 
Their mass eigenstates are defined as 
\begin{align}
\begin{pmatrix}
\varphi_H^{} \\
\varphi_A^{}
\end{pmatrix}
=
R(\theta)
\begin{pmatrix}
H_1 \\
H_2
\end{pmatrix},~
R(\theta)\equiv
\begin{pmatrix}
\cos\theta & -\sin\theta \\
\sin\theta & \cos\theta
\end{pmatrix}. \label{eq:cp-mix}
\end{align}

Now, let us discuss the cross section for $f\bar{f} \to Z^* \to H_1H_2$. 
The $H_1H_2 Z^\mu$ vertex is given by 
\begin{align}
|D_\mu \varphi|^2 &\supset
g_Z^{}Y_\varphi (H_1\overleftrightarrow{\partial}_\mu H_2) Z^\mu, 
\end{align}
where $A\overleftrightarrow{\partial}_\mu B \equiv A(\partial_\mu B) - (\partial_\mu A)B$ and $g_Z^{} \equiv  g/\cos\theta_W$ with $g$ and $\theta_W$ being the $SU(2)_I$ gauge coupling and the weak mixing angle, respectively. 
It is clear that $Y_\varphi \neq 0$ is required to obtain the non-vanishing interaction, and this suggests that $\varphi$ should be an isospin non-singlet field. 
The cross section is then expressed at leading order as 
\begin{align}
\hat{\sigma} = \frac{16\pi \alpha_{\rm em}^2 Y_\varphi^2}{3N_f^c s  \sin^42\theta_W}\frac{v_f^2 + a_f^2}{(1 - \frac{m_Z^2}{s})^2}\lambda^{3/2}\left(\frac{m_{H_1}^2}{s},\frac{m_{H_2}^2}{s}\right), \label{eq:cross}
\end{align}
where $s$ is the squared center-of-mass energy, 
$N_f^c = 3 (1)$ for $f$ being quarks (leptons), and 
$v_f = I_f/2 - Q_f\sin^2\theta_W$ and $a_f=I_f/2$
with $I_f$ being the third component of the isospin of a fermion $f$.
The phase space function is given by $\lambda(x,y) = (1-x-y)^2 - 4xy$. 
The cross section for the four-photon process (\ref{eq:4photon}) is then estimated by $\hat{\sigma} \times \xi$ with 
\begin{align}
\xi \equiv \text{BR}(H_1 \to \gamma\gamma)\times \text{BR}(H_2 \to \gamma\gamma). \label{eq:xi}
\end{align}
For $f$ to be quarks, the cross section should be written as $\sigma \times \xi$ with $\sigma$ being the 
hadronic production cross section for $pp \to H_1H_2$.  

Next, we discuss the decays of $H_{1,2}$. 
We introduce the following Yukawa and scalar interactions: 
\begin{align}
{\cal L}_{\rm int} = -\frac{\sqrt{2}m_f}{v}\hat{\zeta}_f\bar{f}_L f_R\varphi
-\sum_{\alpha}\mu_\alpha S_\alpha S_\alpha^*\varphi + \text{h.c.}, 
\end{align}
where $S_\alpha$ are charged scalars with the electric charge $Q_\alpha$. 
We here do not specify the other properties of $S_\alpha$ such as the isospin. 
These interaction terms can be rewritten in the basis of $H_a$ as 
\begin{align}
{\cal L}_{\rm int} = -\sum_{a = 1,2}\left[\frac{m_f}{v}\bar{f}(\kappa_a^f + i\gamma_5 \tilde{\kappa}_a^f) f
+\sum_\alpha\mu_{\alpha a}S_\alpha S_\alpha^* \right]H_a, \label{eq:int}
\end{align}
with 
\begin{align}
&
\begin{pmatrix}
\kappa_1^f \\
\tilde{\kappa}_1^f 
\end{pmatrix}
=R(\theta)
\begin{bmatrix}
\Re(\hat{\zeta}_f) \\
\Im(\hat{\zeta}_f)
\end{bmatrix},~
\begin{pmatrix}
\kappa_2^f \\
\tilde{\kappa}_2^f 
\end{pmatrix}
=R(\theta)
\begin{bmatrix}
-\Im(\hat{\zeta}_f)\\
\Re(\hat{\zeta}_f) 
\end{bmatrix}, \notag\\
&
\begin{pmatrix}
\mu_{\alpha 1} \\ 
\mu_{\alpha 2}
\end{pmatrix}
= \sqrt{2}R^T(\theta)
\begin{bmatrix}
\Re (\mu_\alpha) \\ 
-\Im (\mu_\alpha)
\end{bmatrix}. \label{eq:couplings}
\end{align}
When the Higgs alignment condition is exactly imposed, $H_a$ do not decay into a weak boson pair, 
while they can decay into a fermion pair and/or a lighter additional Higgs boson associated with a (off-shell) weak boson at tree level. 
At one-loop level, $H_a$ can decay into $\gamma\gamma$, $Z\gamma$ and $gg$.
In order to discuss how the di-photon decay can be important, we give the decay rates into $f\bar{f}$, $gg$
and $\gamma\gamma$ as follows
\begin{align}
\Gamma(H_a \to f\bar{f})&=
\frac{N_f^c m_{H_a}^3}{32\pi v^2} \tau_a^f (|\kappa_a^f|^2 - \tau_a^f[\Re(\kappa_a^f)]^2)\notag\\
&\times\sqrt{1 - \tau_a^f}, \\
 \Gamma(H_a \to gg)&=
 \frac{\alpha_s^2 m_{H_a}^3}{128\pi^3v^2}\notag\\
 &\hspace{-18mm}\times\Bigg\{\Bigg|\sum_f \kappa_a^f I_{\frac{1}{2}}(\tau_a^f)  \Bigg|^2    + \Bigg|\sum_f \tilde{\kappa}_a^f \tilde{I}_{\frac{1}{2}}(\tau_a^f) \Bigg|^2 \Bigg\}, \\
\Gamma(H_a \to \gamma\gamma)&=
\frac{\alpha_{\rm em}^2 m_{H_a}^3}{256\pi^3v^2}\notag\\
&\hspace{-18mm}\times \Bigg\{\Bigg|\sum_fQ_f^2N_f^c\kappa_a^f I_{\frac{1}{2}}(\tau_a^f) 
+\sum_\alpha Q_\alpha^2\frac{v\mu_{\alpha a}}{m_{H_a}^2}I_0(\tau_a^{S_\alpha}) \Bigg|^2 \notag\\
& + \Bigg|\sum_fQ_f^2N_f^c \tilde{\kappa}_a^f \tilde{I}_{\frac{1}{2}}(\tau_a^f) \Bigg|^2 \Bigg\}, \label{eq:gamgam}
\end{align}
where $\tau_a^X = 4m_X^2/m_{H_a}^2$. 
The loop functions are given by~\cite{Gunion:1989we} 
\begin{align}
\begin{split}
I_0(x) &= 2[1-xf(x)], \\
I_{\frac{1}{2}}(x) &= 2x[(x-1)f(x)-1],~~\tilde{I}_{\frac{1}{2}}(x) = 2xf(x),
\end{split}
\end{align}
with 
\begin{align}
f(x) = 
\begin{cases}
\arcsin^2\sqrt{x^{-1}}~~(x\geq 1),\\
-\frac{1}{4}\left[\ln\frac{1+\sqrt{1-x}}{1-\sqrt{1-x}} -i\pi\right]^2~~(x< 1). 
\end{cases}
\end{align}
In Eq.~(\ref{eq:gamgam}), the contribution from the W boson loop is neglected, because of the Higgs alignment condition. 
We note that the decay rates of $H_a \to Z\gamma$ can be comparable with those of $H_a \to \gamma\gamma$ as long as $m_{H_a} \gg m_Z^{}$, which will be included in our numerical analysis given below.  

In the CP-conserving (CPC) limit, i.e., $\theta \to 0$ and $\Im(\hat{\zeta}_f) = \Im(\mu_i) = 0$, $H_1$ ($H_2$) behaves as a CP-even (CP-odd) scalar boson, and 
the $S_\alpha^\pm$ loop contribution to $H_2 \to \gamma\gamma$ vanishes. 
In this case, $\text{BR}(H_2 \to \gamma\gamma)$ cannot be significant due to the dominant  $H_2 \to f\bar{f}/gg$ modes. 
In fact, when we consider only the top-loop contribution to the $H_2 \to \gamma\gamma/gg$ modes, 
the ratio $\Gamma(H_2 \to \gamma\gamma)/\Gamma(H_2 \to gg)$ is given by $(\alpha_{\rm em}N_t^cQ_t^2/\sqrt{2}\alpha_s)^2 \simeq 4\times 10^{-3}$. 
Thus, $H_2 \to \gamma\gamma$ cannot be the dominant mode. 
For $m_{H_2} \geq 2m_t$, $\text{BR}(H_2 \to \gamma\gamma)$ is even more suppressed by the $H_2 \to t\bar{t}$ mode. 
On the other hand, for the case with CPV, the $S_\alpha^\pm$-loop contributes to the $H_2 \to \gamma\gamma$ mode, so that $\text{BR}(H_2 \to \gamma\gamma)$ can be large. 
In particular, if both $\mu_{\alpha 1}$ and  $\mu_{\alpha 2}$ are relatively larger than the $\hat{\zeta}_f$ parameters, {\it both }
the branching ratios of $H_{1,2} \to \gamma\gamma$ can be sizable. 
Therefore, a larger value of $\xi$ defined in Eq.~(\ref{eq:xi}) can be a telltale sign of CPV in the Higgs sector. 

{\it Concrete Models --} Let us discuss the four-photon process (\ref{eq:4photon}) in the general two Higgs doublet model (2HDM) without imposing any additional symmetries as a prototype of an extended Higgs sector. 
The scalar multiplet $\varphi$ is then identified with another isospin doublet field $\Phi'$ with $Y_{\Phi'} = 1/2$. 
We can take $\langle \Phi' \rangle = 0$ without loss of generality, because $\Phi'$ can be regarded as the field defined in the Higgs basis~\cite{Davidson:2005cw}. 

The most general Higgs potential is written as
\begin{align}
V &= m^2|\Phi|^2 + M^2|\Phi'|^2 - (\mu^2\Phi^\dagger \Phi' + \text{h.c.}) \notag\\
& + \frac{\lambda_1}{2}|\Phi|^4 +  \frac{\lambda_2}{2}|\Phi'|^4 + \lambda_3|\Phi|^2|\Phi'|^2 + \lambda_4|\Phi^\dagger \Phi'| \notag\\
& + \left(\frac{\lambda_5}{2}\Phi^\dagger \Phi' + \lambda_6|\Phi|^2 + \lambda_7|\Phi'|^2 \right)(\Phi^\dagger \Phi')+ \text{h.c.}, \label{eq:pot}
\end{align}
where $\mu^2$ and $\lambda_{5,6,7}$ are generally complex parameters.  

%
Imposing the stationary conditions, we can eliminate the parameters $m^2$ and $\mu^2$. 
The mass matrix for the neutral Higgs bosons are then given in the basis of ($\sqrt{2}\Re \Phi^0,\sqrt{2}\Re \Phi^{\prime 0},\sqrt{2}\Im \Phi^{\prime 0}$) as 
\begin{align}
v^2
    \begin{pmatrix}
      \lambda_1	& \Re \lambda_6 & -\Im \lambda_6 \\
      \Re \lambda_6 &\frac{M^2}{v^2} + \frac{\lambda_3+\lambda_4+\Re \lambda_5}{2}&-\frac{\Im \lambda_5}{2} \\
      -\Im \lambda_6 &-\frac{\Im \lambda_5}{2} &\frac{M^2}{v^2}+\frac{\lambda_3+\lambda_4-\Re \lambda_5}{2}
    \end{pmatrix}. 
    \label{eq:massmatrix}
    \end{align}
We can remove the phase of $\lambda_5$ by the field redefinition without loss of generality. 
The Higgs alignment condition is given by 
\begin{align}
\lambda_6 = 0, \label{eq:higgs-alignment}
\end{align}
in which the mass matrix takes the diagonal form, i.e., $\theta = 0$ in Eq.~(\ref{eq:cp-mix}). 
The basis invariant form of the CPV quantities in the 2HDM has been found in Ref.~\cite{Lavoura:1994fv} as follows: 
\begin{align}
J_1 \propto \Im [\lambda_5^*\lambda_6^2],~~
J_2 \propto \Im [\lambda_5^*\lambda_7^2],~~
J_3 \propto \Im [\lambda_6^*\lambda_7], 
\end{align}
where CP-symmetry is broken if at least one of the three invariants is non-zero. 
Therefore, our scenario $\lambda_6 = \Im \lambda_5 = 0$ with $\Im \lambda_7\neq 0$ gives $J_2 \neq 0$, and we definitely have CPV in the potential. 
Under $\lambda_6 = \Im \lambda_5 = 0$, $J_2$ can also be written as 
\begin{align}
J_2\propto  (m_{H_1}^2 - m_{H_2}^2)\Im [\lambda_7^2]. 
\end{align}
This suggests that the phase of $\lambda_7$ turns out to be unphysical when two masses are degenerate, i.e., $m_{H_1} = m_{H_2}$. 
In the 2HDM, $S_\alpha^\pm$ are identified with the singly-charged Higgs bosons $\Phi^{\prime\pm} (\equiv H^\pm)$, and the scalar coupling $\mu_\alpha$ defined in Eq.~(\ref{eq:int})
is expressed as $\mu_\alpha = v\lambda_7/\sqrt{2}$. 

The Yukawa interactions are generally given in the mass basis for fermions as 
\begin{align}
&{\cal L}_Y =
-\frac{\sqrt{2}}{v}\Big[\bar{Q}^u_LM_u i\tau_2\left(\Phi^* + \rho_u \Phi^{\prime *}\right)u_R \notag\\
&+\bar{Q}^d_LM_d\left(\Phi + \rho_d\Phi'\right)d_R 
+\bar{L}_LM_e\left(\Phi + \rho_e \Phi'\right) e_R \Big]+\text{h.c.}, 
\end{align}
where $Q_L^d = (V^\dagger u_L,d_L)^T$ and $Q_L^u = (u_L,Vd_L)^T$ with $V$ being the Cabibbo-Kobayashi-Maskawa matrix. 
In the above expression, $M_f$ ($f=u,d,e$) are the diagonalized mass matrix, and $\rho_f$ are general complex $3\times 3$ matrices. 
In order to avoid flavor-changing neutral currents via Higgs boson mediations at tree level, 
we impose the so-called Yukawa alignment~\cite{Pich:2009sp}, i.e., 
\begin{align}
\rho_f = \zeta_f I_{3\times 3}~~(f= u,d,e), 
\end{align}
where $\zeta_f$ are complex parameters and $I_{3\times 3}$ is the $3\times 3$ unit matrix.
Comparing Eq.~(\ref{eq:int}), we can identify $\hat{\zeta}_f = \zeta_f$. 


 \begin{figure*}[t]
 \begin{center}
  \includegraphics[width=55mm]{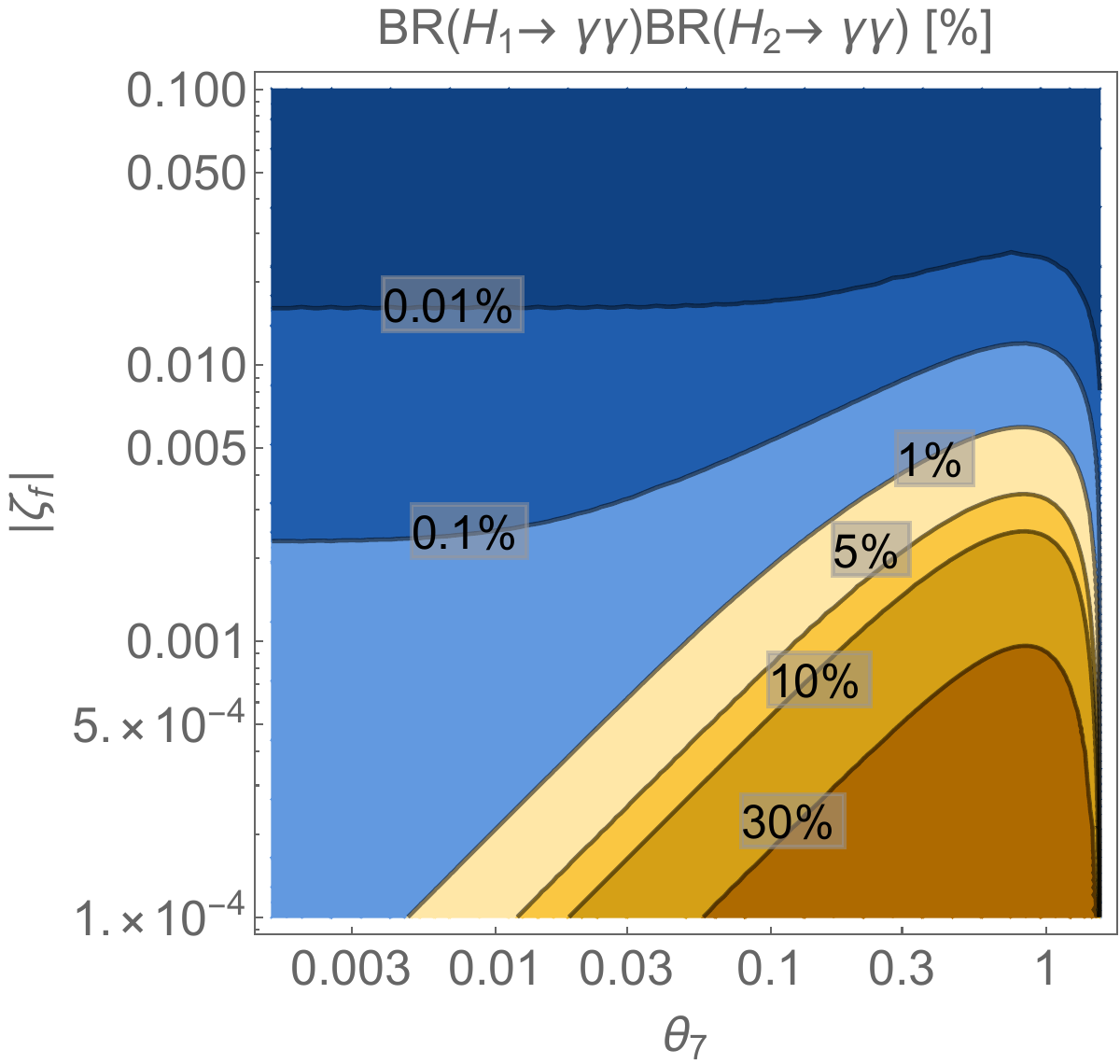}
  \includegraphics[width=55mm]{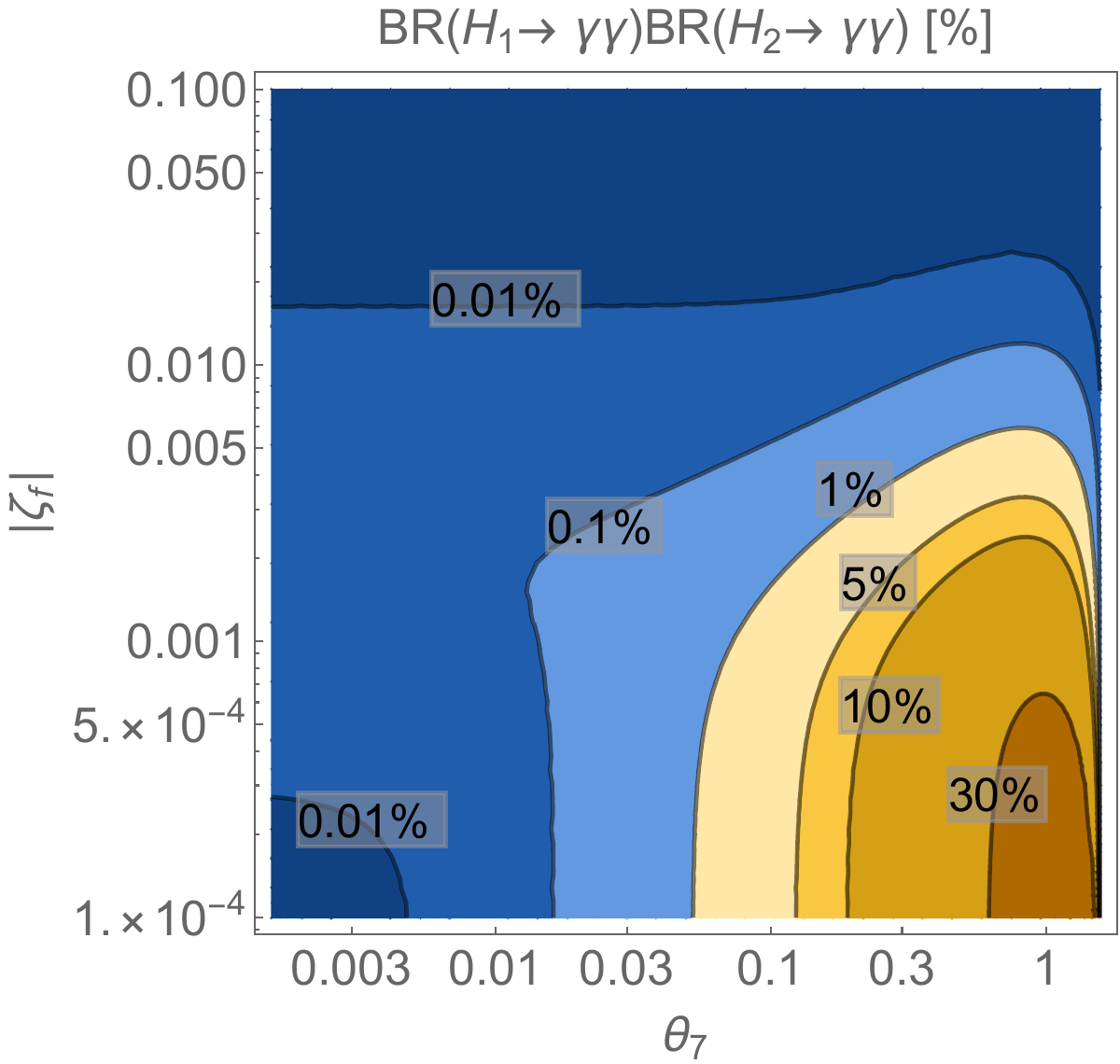}
  \includegraphics[width=55mm]{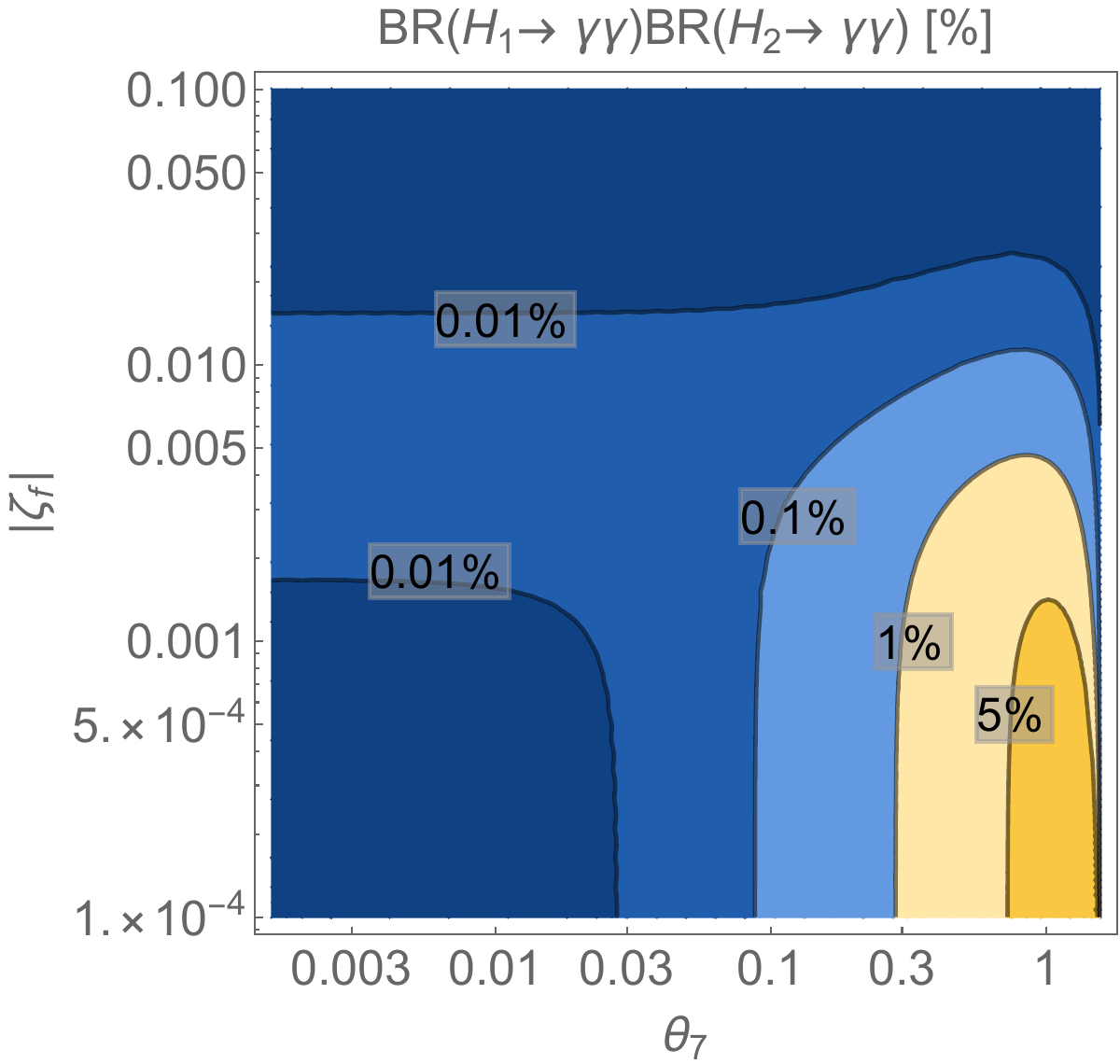}
    \caption{Contour plots of the product of the branching ratios $\xi = \text{BR}(H_1 \to \gamma\gamma) \times \text{BR}(H_2 \to \gamma\gamma)$ 
on the $\theta_7$-$|\zeta_f|$ plane in the 2HDM. 
The left, center and right panels show the case with $m_{H_2} - m_{H_1}$ to be 0.1, 5 and 10 GeV, respectively. 
For all the plots, we take $\zeta_f = \zeta_u = \zeta_d = \zeta_e$, $\theta_f = 0$, $m_{H_1} =m_{H^\pm}= 250$ GeV and $|\lambda_7| = 1$. }
    \label{fig:brbr}
 \end{center}
 \end{figure*}

 \begin{figure}[t]
 \begin{center}
  \includegraphics[width=70mm]{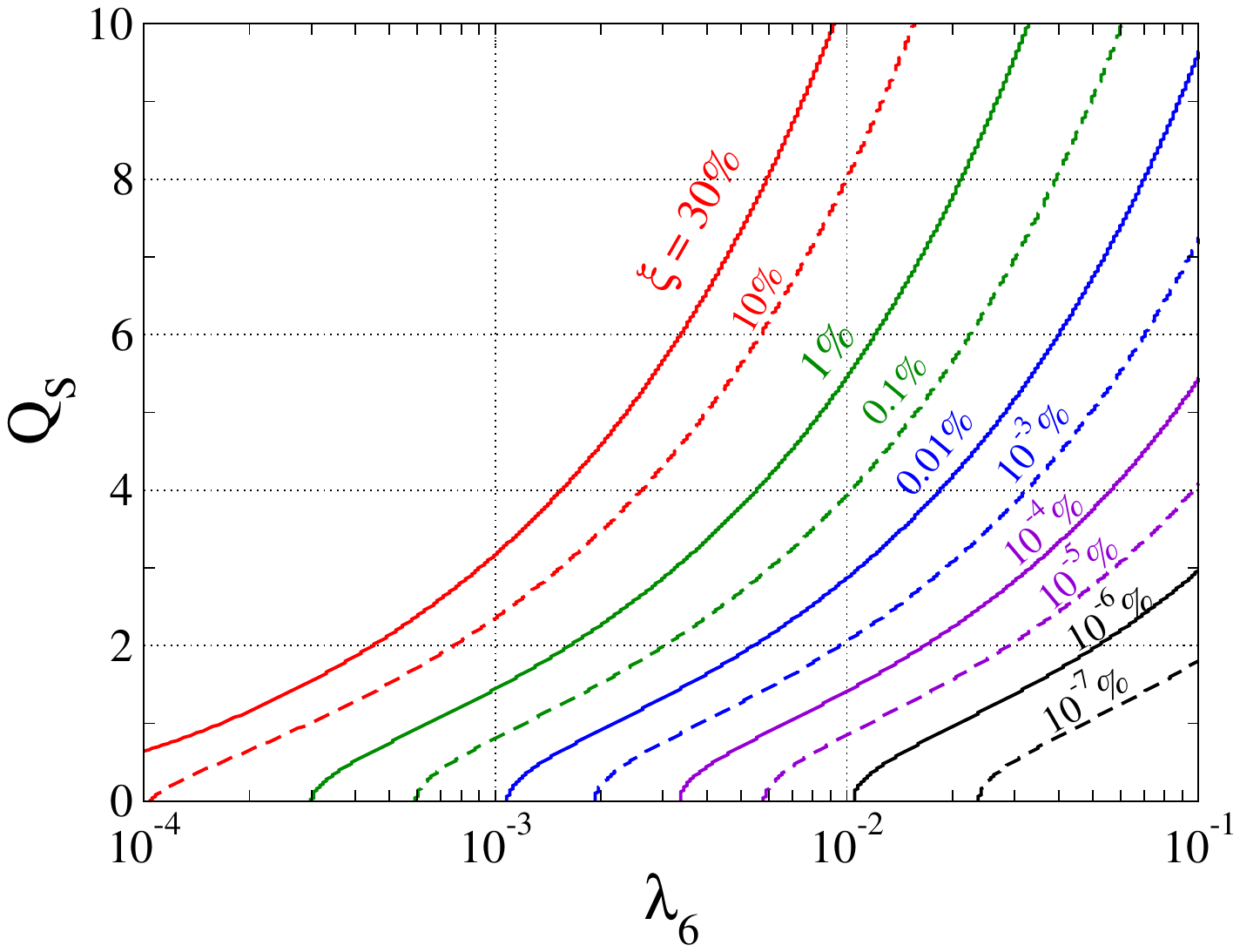}
    \caption{Contour plots of $\xi = \text{BR}(H_1 \to \gamma\gamma) \times \text{BR}(H_2 \to \gamma\gamma)$ 
on the $\lambda_6$-$Q_S$ in the 2HDM with an additional charged scalar with the electric charge $Q_S$. 
We take $\zeta_f = \zeta_u = \zeta_d = \zeta_e$, $\theta_f = 0$, $m_{H_2} - m_{H_1}=5$ GeV, $m_{H_1} =m_{H^\pm} = M = 250$ GeV, $|\lambda_7| = 1$ and $\theta_7 = \pi/4$.
}
    \label{fig:brbr2}
 \end{center}
 \end{figure}


In Fig.~\ref{fig:brbr}, we show the contour of $\xi$ as a function of the phase $\theta_7 \equiv \text{arg}(\lambda_7)$ and the magnitude of $\zeta_f$ in the 2HDM. We take the mass difference $\Delta m \equiv m_{H_2} - m_{H_1}$ to be 0.1 GeV (left), 5 GeV (center) and 10 GeV (right). 
As expected, in the CPC limit $\theta_7 \to 0$ 
the value of $\xi$ is given to be of order 0.1\% or smaller, because $H_2 \to \gamma\gamma$ cannot be significant. 
On the other hand, $\xi$ takes larger values especially in the lower-right region, i.e., larger $\theta_7$ and/or smaller $|\zeta_f|$. 
It is also seen that a larger value of $\xi$ is realized for smaller $\Delta m$, because the decay mode $H_2 \to H_1Z^*$ is phase space suppressed. 

Let us comment on how the branching ratio is modified if we consider the case with a slight deviation from the Higgs alignment limit, i.e.,  $\lambda_6 \neq 0$. 
In this case, $H_a \to WW/ZZ/Zh$ can be important, by which the branching ratio into di-photon is highly suppressed. 
As shown in Fig.~\ref{fig:brbr2}, $\xi$ takes about $10\%$ and $10^{-2}\%$ for $\lambda_6 = 10^{-4}$ and $10^{-3}$, respectively, in the 2HDM (case with $Q_S= 0$). 
Such a quite sensitive dependence on $\lambda_6$ can be milder if additional charged scalars are present. 
For instance, if we introduce a charged singlet scalar with the electric charge $Q_S$ and the same mass and $\mu_{\alpha a}$ as those of the charged Higgs boson $H^\pm$, then 
$\xi\simeq 1\%$ can be obtained for $(Q_{S},|\lambda_6|)$ to be around e.g., (1,$10^{-3}$) and (5,$10^{-2}$). 
For the masses of $H_a$ below $2m_W$, larger values of $|\lambda_6|$ are possible while keeping $\xi$ to be sizable due to the phase space suppression of the $WW/ZZ/Zh$ modes.

{\it Four-photon process at LHC} -- We discuss how large cross section of the four-photon process (\ref{eq:4photon}) can be obtained at LHC in the 2HDM with the Higgs alignment. 

We first discuss existing experimental constraints on the parameter space in the 2HDM with the Higgs alignment. 
We take into account the constraints from the eEDM experiments, $|d_e| < 4.1\times 10^{-30}e\,\text{cm}$ (90\% CL)~\cite{Roussy:2022cmp}. 
We confirmed that the constraints from the other EDMs such as the neutron EDM~\cite{Abel:2020pzs} do not further exclude the parameter space allowed by the eEDM. 
In addition, we impose the following two constraints coming from LHC: 
(A) searches for a di-photon resonance~\cite{ATLAS:2021uiz} and (B) those for multi-photon ($\geq 3 \gamma$) final states~\cite{ATLAS:2015rsn}. 
For (A), we consider the gluon fusion (ggF) $gg \to H_a$~\cite{Georgi:1977gs} and the EW $q\bar{q}' \to H^\pm H_a$~\cite{Cao:2003tr} production processes. 
We estimate the production cross section for ggF using \texttt{SusHi}~\cite{Harlander:2012pb,Harlander:2016hcx} at NNLO in QCD. 
Since the cross section for ggF is proportional to $|\zeta_u|^2$, we find that the limit coming from ggF is negligible 
for our chosen parameter space,  $|\zeta_f|\ll 1$.
However, the EW production remains crucial and deliver a stringent limit on the parameter space. 
For (B), the EW production $H_1H_2$ with their $\gamma\gamma$ and/or $Z\gamma$ decays can give rise to the multi-photon signal. 
In what follows, we consider the case with $m_{H_1} = m_{H^\pm}$ and $m_{H_2} > m_{H_1}$, so that 
the decay $H_2 \to H_1Z^*$ provides $H_1H_1Z^*$ in the intermediate state, and it can also contribute to the multi-photon signal.


 \begin{figure}[t]
 \begin{center}
  \hspace{-5mm}\includegraphics[width=45mm]{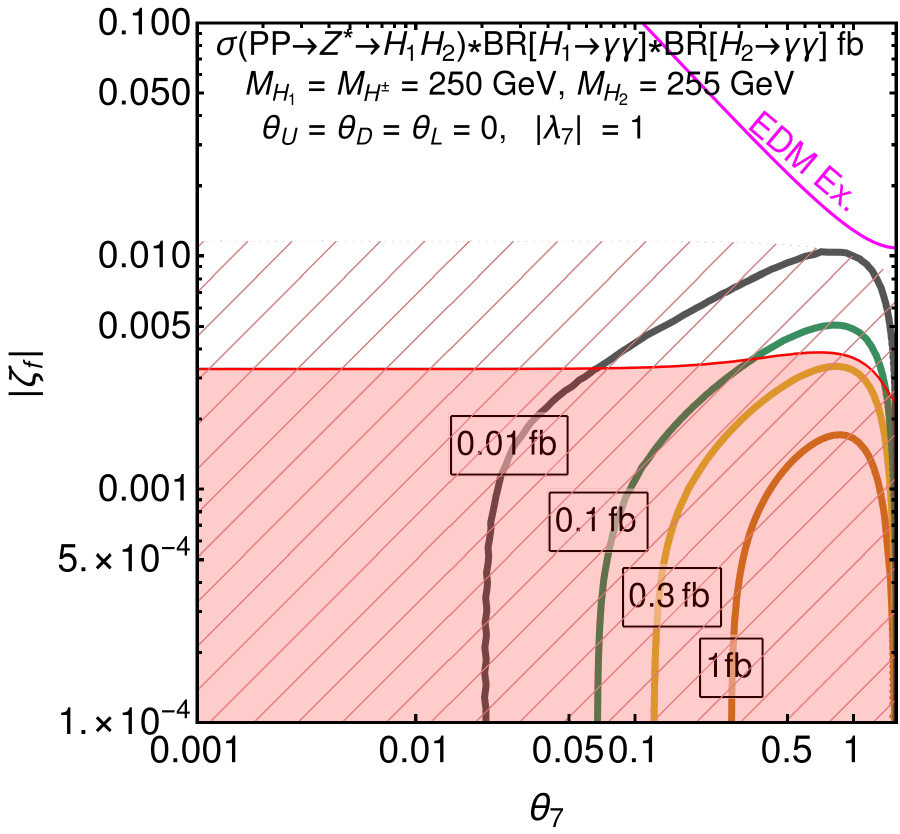}
  \includegraphics[width=45mm]{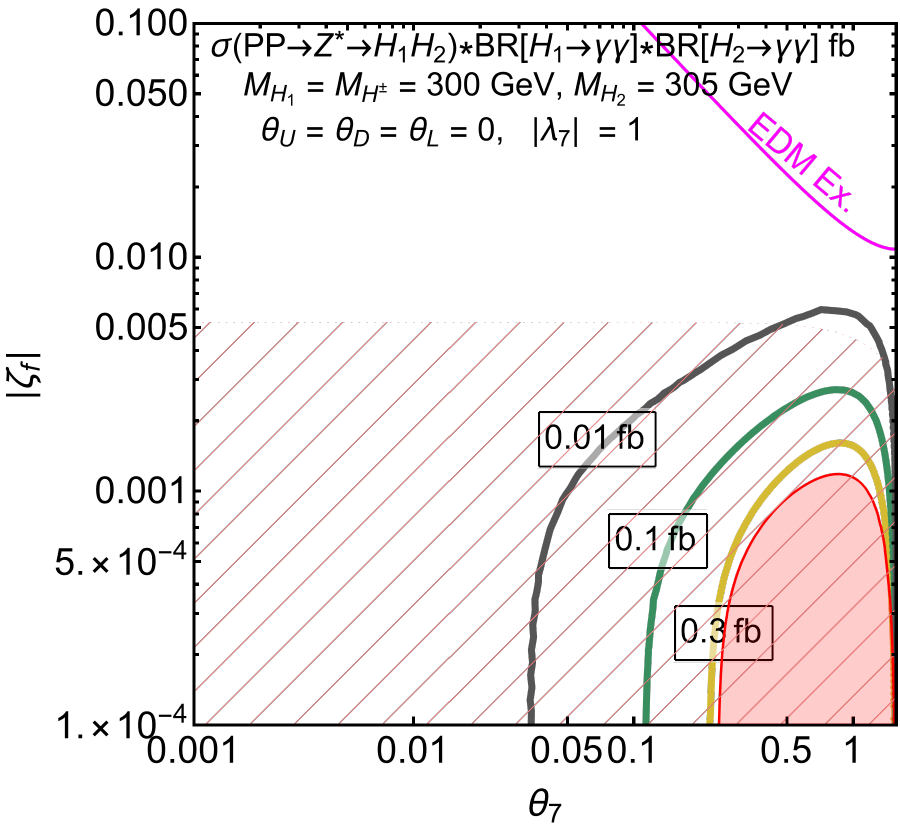}
  \caption{Contour plots of the cross section for the four-photon process given in (\ref{eq:4photon}) at LHC on the $\theta_7$-$|\zeta_f|$ plane in the 2HDM. 
The left and right panels show the case with $m_{H_1} = m_{H^\pm} = 250$ and 300 GeV, respectively. 
For all the plots, we take $m_{H_2} - m_{H_1} = 5$ GeV, $|\lambda_7| = 1$ and $\theta_f = 0$. 
The red shaded region is excluded by the constraint from the di-photon searches, while the region above the black solid curve
is excluded by the constraint from the eEDM. 
The hashed region is expected to be explored with more than 2$\sigma$ level from di-photon searches at the HL-LHC with 3 $\text{ab}^{-1}$. }
    \label{fig:LHC}
 \end{center}
 \end{figure}

Fig.~\ref{fig:LHC} shows one of the main results of our analysis in the 2HDM for $m_{H_1} = 250$ GeV (left panel) and  300 GeV (right panel)
with the mass difference $m_{H_2} - m_{H_1}$ of 5 GeV. 
This figure shows the contour of the cross section for the four-photon final state under all the experimental constraints explained above in the 2HDM. 
We see that the severe bound exists from the di-photon search [constraint (A)] indicated by the red shaded region. 
On the other hand, the limit coming from the multi-photon searches (B) do not appear in these figures, 
because only the Run-I data with at 8 TeV and 20 fb$^{-1}$ are available, which are rather weak to exclude the parameter region shown here. 
The constraint from the eEDM excludes the region with larger $|\zeta_f|$ and/or larger $\theta_7$. 
We also show the region expected to be explored with more than 2$\sigma$ level at the high-luminosity LHC (HL-LHC) by the hashed region, which is obtained by extrapolating 
the current result of the di-photon search~\cite{ATLAS:2021uiz}.
We see that the searches for EDM and the di-photon resonance at LHC play complementary roles to each other. 
Before concluding, we would like to emphasize that the di-photon signal, although far-reaching, 
does not specify the CP nature of the scalar potential. Searching for the proposed four-photon signal is essentially important to probe CPV in the extended Higgs sector at LHC. 
 

%
%

{\it Discussions and Conclusions --} Let us first comment on the other possibility for the multiplet $\varphi$ than the isospin doublet. 
As mentioned above, $\varphi$ cannot be an isospin singlet because of $Y_\varphi = 0$. 
For an isospin triplet, we can consider the one with $Y_\varphi = 1$, 
but this model does not contain a physical CP-violating phase in the potential~\cite{Ferreira:2021bdj}. Thus, $\xi$ cannot be large. 
The same thing holds for models with $\varphi$ whose isospin is larger than triplet except for the case with $\varphi$ to be quadruplet with $Y_\varphi = 1/2$. 
For the latter, the potential contains two terms $(\Phi\varphi^*)^2$ and 
$(\Phi\Phi^*\Phi^*\varphi)$, and one of the phases for these couplings can be physical, so that a large $\xi$ value can be realized. 
For models with more than one extra scalar fields, e.g., a model with two triplets~\cite{Ferreira:2021bdj,Chen:2023ins}, 
physical CP-violating phases can appear in the potential, and a larger value of $\xi$ can be realized. 

We also comment on four-photon final states realized in the other scenarios. 
In the CPC Type-I 2HDM, 
one can consider the sizable cross section for the exact four-photon final state via $gg \to h \to H_1H_1 \to 4\gamma$~\cite{Arhrib:2017uon}.
%
There are two crucial differences between the above process and that in (\ref{eq:4photon}), i.e., 
(i) the invariant mass distribution for the di-photon system shows only one peak at $m_{H_1}$ (after taking into account combinatorics of four photons) in the above but two peaks at $m_{H_1}$ and $m_{H_2}$ in our process
and (ii) a deviation from the Higgs alignment is required to obtain the $h \to H_1H_1$ decay in the above process. 
We also note that the 2HDMs with a softly-broken $Z_2$ symmetry, including the Type-I 2HDM, can provide a non-zero CP-violating phase in the potential, while this phase introduces a mixing among three neutral Higgs bosons. 
Therefore, such 2HDMs with CPV may be able to give a larger value of $\xi$, but they also introduce a larger deviation in the couplings of $h$ from the SM values.

To conclude, our proposed scenario provides a sizable number of four-photon events coming from the di-photon decay of two additional neutral Higgs bosons 
when the $|\lambda_7|$ ($|\zeta_f|$) parameter is taken to be larger (smaller) with order one $\theta_7$. 
The key point of the four-photon process realized in our scenario is the appearance of two distinguishable peaks in the invariant mass of the di-photon system and the 
compatibility with the Higgs alignment limit, which cannot be realized in the 2HDMs with the softly-broken $Z_2$ symmetry. 
We advocate that, in addition to di-photon processes, 
it is worthwhile to systematically investigate the multi-photon process at LHC. As we have shown, such a process can be crucial to identify CPV in the Higgs sector.


{\it Acknowledgments --}  This work was supported in part by JSPS KAKENHI Grants Nos. 20H00160, 22F21324 and 23K17691.

\bibliography{references}


\end{document}